\documentclass[useAMS,usenatbib,usegraphicx]{mn2e}
\bibpunct[,]{(}{)}{;}{a}{}{}

\usepackage{amssymb}

\newcommand{\hm}{h^{-1}}

\title[Mass Segregation in Groups and Clusters]
{Semi-Analytic Model Predictions of Mass Segregation from Groups to Clusters}
\author[E.~Contini et al.]
        {E.~Contini, \thanks{Email: contini@pmo.ac.cn} 
         X.~Kang,    \thanks{Email: kangxi@pmo.ac.cn}
         \\        
         Purple Mountain Observatory, the Partner Group of MPI f\"{u}r Astronomie, 2 West Beijing Road, Nanjing 210008, China \\}

\pagerange{\pageref{firstpage}--\pageref{lastpage}}
\pubyear{2015}

\begin{document}

\maketitle

\label{firstpage}

\begin{abstract} 
Taking advantage of a high-resolution simulation coupled with a state-of-art semi-analytic model of galaxy
formation, we probe the mass segregation of galaxies in groups and clusters, focusing on which physical mechanisms 
are driving it. We find evidence of mass segregation in groups and clusters up to the virial radius, both looking 
at the galaxy stellar mass and subhalo mass. The physical mechanism responsible for that is consistent with dynamical 
friction, a drag-force that brings more massive galaxies faster towards the innermost regions of the halo. At odds 
with observational results, we do not find the inclusion of low-mass galaxies in the samples, down to stellar mass 
$M_* = 10^9 \, M_{\odot}$, to change the overall trend shown by intermediate and massive galaxies. Moreover, stellar 
stripping as well as the growth of galaxies after their accretion, do not contribute either in shaping mass segregation 
or mixing the radial mass distribution. Beyond the virial radius we find an ``anti-mass segregation" in groups that 
progressively weakens in clusters. The continuous accretion of new objects and recent merger events play a different 
role depending on the halo mass onto which accreting material is falling.  
\end{abstract}


%%%%%%%%%%%%%%%%%%%%%%%%%%%%%%%%%%%%%%%%%%%%%%%%%%%%%%%%%%%%%%%%%%%%%%%%%%%%%%%%
\section[]{Introduction} 
\label{sec:intro}
The current scenario of galaxy formation can be divided in two distinct galaxy evolutionary families: the \emph{nature} scenario,
where galaxy properties depend on the state of the galaxies at the time of their formation, and the \emph{nurture} scenario, where 
the galaxy properties are instead the product of environmentally driven processes that take place after the galaxy is accreted 
on large systems, such as groups and clusters. As highlighted by \cite{gab12a}, galaxies experience several different 
kinds of environment during their lifetime, thus leading to the conclusion that the aforementioned scenarios are twisted 
together. The mass distribution of galaxies in different environments is therefore the result of many physical processes that 
operate during galaxy evolution, such as gas cooling, star formation, stripping and mergers, and the environment is believed 
to have a key-role in shaping galaxy properties (e.g., \citealt{kauffmann04,tanaka04,blanton05,weinmann06,blantonberlind07}). 
We now know that galaxy properties such as mass, colour, gas content and age are strongly related to the environment, such that
galaxies in denser environments are typically more massive, redder, less gas-rich and older.

Dynamical friction is a key factor in the link between galaxy growth and environment. This has been pointed-out by several authors 
in the context of dark matter substructures (see e.g. \citealt{delucia, myself}) as well as the connection between galaxy stellar mass 
and subhalo mass (e.g., \citealt{gao3,wang,vale}). According to the dynamical friction formula (given by \citealt{chandrasekhar43}), 
more massive galaxies fall faster towards the innermost region of the cluster, thus leading to a segregation in mass. Under this picture,
the characteristic galaxy mass is expected to increase with increasing halo mass (\citealt{yang05,zheng05,vandenbosch08}) because 
galaxies are more massive in larger objects, and more massive galaxies should preferentially reside in the centre. Hence, if dynamical 
friction plays an important role, its effect on the distribution of galaxies in large system must be observable. On the other hand, 
stellar stripping due to tidal forces between galaxies and the cluster potential, as well as ongoing star formation in galaxies as 
they fall towards the center, might affect the distribution in mass. Tidal stripping removes part of the stellar mass of galaxies, and 
is expected to be more effective in massive galaxies (\citealt{rudick09,martel12,watsonconroy13,laporte13,myself2}), while star 
formation leads to an increase in stellar mass.

Despite the large amount of observations and effort spent on the topic, there is not yet a common agreement in the literature. There
is evidence in favour of mass segregation in large systems (\citealt{lares04,mcintosh05,vandenbosch08b,presotto12,balogh14,roberts15}), 
but at the same time there is significant evidence for a lack of mass segregation in clusters (\citealt{vonderlinden10,vulcani13,ziparo13}). 
Thus, the controversy over mass segregation remains unsolved. 

Recently, \cite{roberts15}, using group catalogues derived from the Sloan Digital Sky Survey Data Release 7 (SDSS DR7, \citealt{sdssDR7}), 
find evidence for mass segregation out to $2\cdot R_{200}$, with strength that scales inversely with halo mass. These authors argue that 
the conflicting results in the literature might be reconciled by considering that mass segregation appears more evident once low-mass 
galaxies are included in the sample, and that it is a function of halo mass, with clusters showing little or no segregation. This points-out 
the need to probe a large range of halo mass and different thresholds in galaxy stellar mass, and the need of timely investigations using 
theoretical models.

Theoretical models have not sufficiently contributed to the debate. Only very recently \cite{vulcani14} (hereafter V14), by means of 
semi-analytic models of galaxy formation (\citealt{dlb} and \citealt{qiguo}), investigate the galaxy stellar mass function as a 
function of the environment. Although they do not explicitly focus on mass segregation, these authors find that the stellar mass 
function does not strongly depend on the cluster-centric distance, although there is a hint of mass segregation in low and intermediate 
halo masses.

In this letter we focus on the role played by dynamical friction on the observed mass segregation in groups and clusters. We 
take advantage of a state-of-art semi-analytic model which includes a refined treatment of stellar stripping that can be switched-on/off.
Our simulation allows to achieve the goals highlighted above, since we can probe a large range of halo mass and use different thresholds
in stellar mass in order to: a) investigate the dependence of mass segregation with halo mass; b) understand if it is driven by the inclusion 
of low-mass galaxies in the sample; c) isolate the role of stellar stripping; d) isolate the growth of galaxies after accretion.

%%%%%%%%%%%%%%%%%%%%%%%%%%%%%%%%%%%%%%%%%%%%%%%%%%%%%%%%%%%%%%%%%%%%%%%%%%%%%%%%
\section[]{Methods}  
\label{sec:methods}
The simulation used in this paper and the semi-analytical model are based on \cite{kang12}. We refer the readers to that 
paper for details. Here we simply introduce the main prescriptions. The simulation was performed using Gadget-2 code 
(\citealt{springel05b}) with cosmological parameters adopted from the WMAP7 data release (\citealt{komatsu}), namely: 
$\Omega_{\lambda}=0.73, \Omega_{m}=0.27, \Omega_{b}=0.044$, $h=0.7$ and $\sigma_{8}=0.81$. The simulation box is $200 \, Mpc/h$ 
on each side using $1024^{3}$ particles, each with mass $5.64\cdot 10^8 \, M_{\odot}h^{-1}$. The merger trees are constructed by 
following the subhaloes resolved in FOF haloes at each snapshot (e.g., \citealt{kang05}) making use of the algorithm SUBFIND 
(\citealt{springel}). The semi-analytic model is then grafted on the merger trees and self-consistently models the physical 
processes governing galaxy formation, such as gas cooling, star formation, supernova and active galactic nucleus feedback. The 
galaxy luminosity and colours are calculated based on the stellar population synthesis model of \cite{bc03} using a Chabrier 
stellar initial mass function (\citealt{chabrier03}). Although we select galaxies with stellar mass larger than 
$10^9 \, M_{\odot}h^{-1}$, it is worth noting that the stellar mass completeness of the galaxy catalogue provided by the model 
run on the simulation is about $10^8 \, M_{\odot}h^{-1}$.

We updated the semi-analytic model described above by adding the prescription \emph{Model Tidal Radius+Mergers} presented in
\cite{myself2}, which accounts for the formation of the intra-cluster light via stripping processes and mergers. The stellar
stripping channel accounts for tidal forces that might strip a given fraction of stellar mass from satellites galaxies as they 
approach the innermost regions of dense environments such as galaxy clusters. Depending on the strength of the tidal force, the 
galaxy might be totally disrupted. The merger channel, instead, considers that a fraction (20 per cent in that model) of the 
satellite stellar mass gets unbound in violent relaxation processes that happen during galaxy-galaxy mergers. In \cite{myself2}
we have verified that such a simple prescriptions reproduces approximately the results of the numerical simulations by
\cite{alvaro}. It is however worth noting that, in reality, the fraction of stars that are unbound should depend on the orbital 
circularity (\citealt{alvaro14}). 

%%%%%%%%%%%%%%%%%%%%%%%%%%%%%%%%%%%%%%%%%%%%%%%%%%%%%%%%%%%%%%%%%%%%%%%%%%%%%%%%%%%%%%%%%%%%%%%%%%%%%%%%%%%%%%%%%%
\section{Results}
\label{sec:results}
In order to examine environmental dependences we study 4 samples of haloes, ranging from small groups (with mass $M_{200}$ in the 
range $[10^{13}-5\cdot10^{13}] M_{\odot} \hm$), to large clusters (with mass larger than $5\cdot10^{14} M_{\odot} \hm$). Details
of the samples are given in Table \ref{tab:tab1}.

\begin{table}
\caption{Our simulated haloes have been split in four samples,
according to the halo mass. In the first column, we give the name
of the sample, the second column indicates the range of
$M_{200}$ values corresponding to each sample, while the third
gives the number of haloes in each sample.}
\begin{center}
\begin{tabular}{llllll}
\hline
Sample & Mass Range ($M_{200}$)& Numb. of Haloes \\
\hline
Small Groups   & $[10^{13}-5\cdot10^{13}] M_{\odot} \hm$  & 2286 \\
Large Groups   & $[5\cdot 10^{13}-10^{14}] M_{\odot} \hm$ & 275  \\
Small Clusters & $[10^{14}-5\cdot10^{14}] M_{\odot} \hm$  & 146  \\
Large Clusters & $>5\cdot10^{14} M_{\odot} \hm$           & 9    \\
\hline
\end{tabular}
\end{center}
\label{tab:tab1}
\end{table}

\begin{figure*} 
\begin{center}
\begin{tabular}{cc}
\includegraphics[scale=.75]{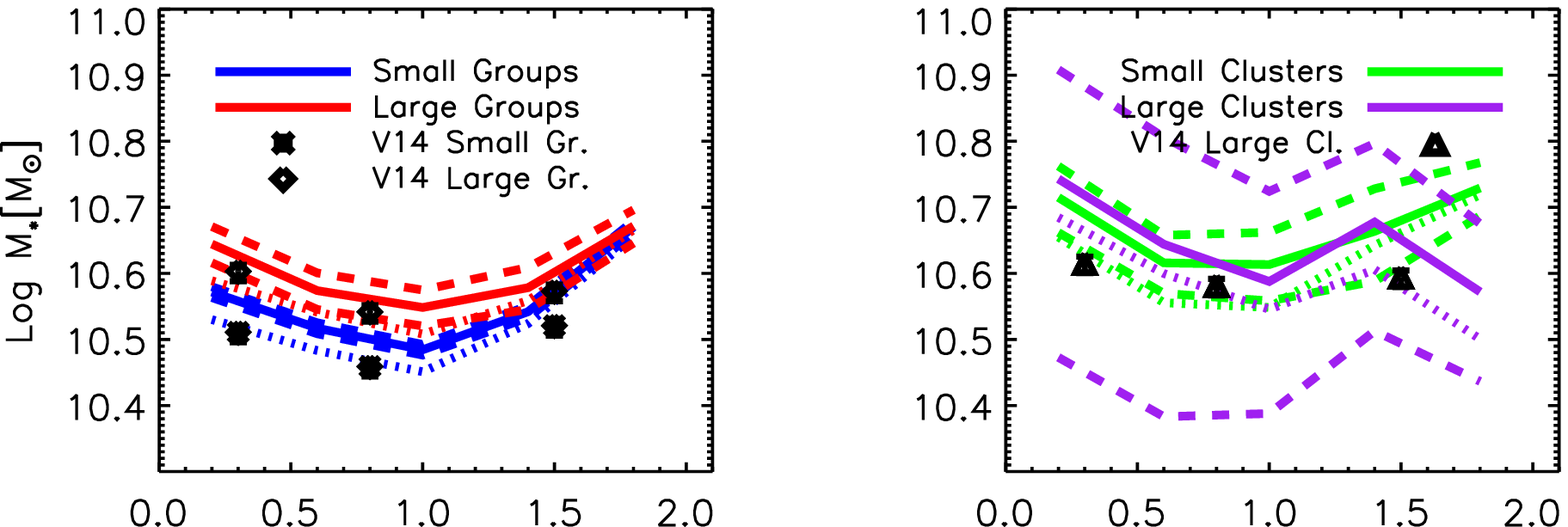} \\ % 0.75
\includegraphics[scale=.75]{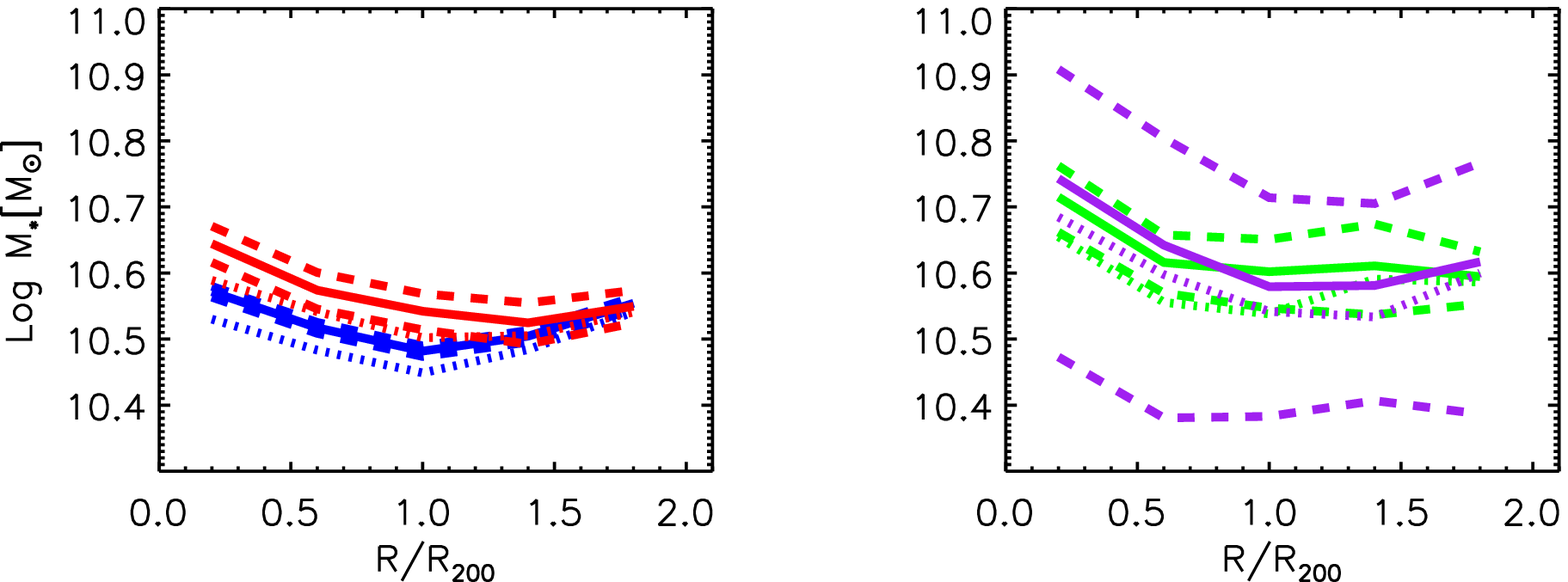} \\ % 0.75
\end{tabular}
\caption{Mean stellar mass of galaxies as a function of radial distance considering the mass at redshift $z=0$
(solid lines) and at the time of accretion (dotted lines). Dashed lines represent 1$\sigma$ statistical errors
computed for the mass at redshift $z=0$ (we omit the scatter around the mean mass at accretion for a more 
readable plot). All galaxies having stellar mass larger than $10^{10} M_{\odot}$ belonging to the FOF-group are 
considered (top panels), for halo masses within $[10^{13}-5\cdot10^{13}] M_{\odot} \hm$ (blue lines), 
$[5\cdot 10^{13}-10^{14}] M_{\odot} \hm$ (red lines), $[10^{14}-5\cdot10^{14}] M_{\odot} \hm$ (green lines), 
and masses $>5\cdot10^{14} M_{\odot} \hm$ (purple lines). Symbols represent data points of \citealt{vulcani14}, 
which are predictions of \citealt{dlb} model. The bottom panels show the same information for the same
samples of haloes, but considering all galaxies  within a sphere of radius $2\cdot R_{200}$ centred on the halo
centre.}
\label{fig:mass_segr}
\end{center}
\end{figure*}

\begin{figure*} 
\begin{center}
\includegraphics[scale=.75]{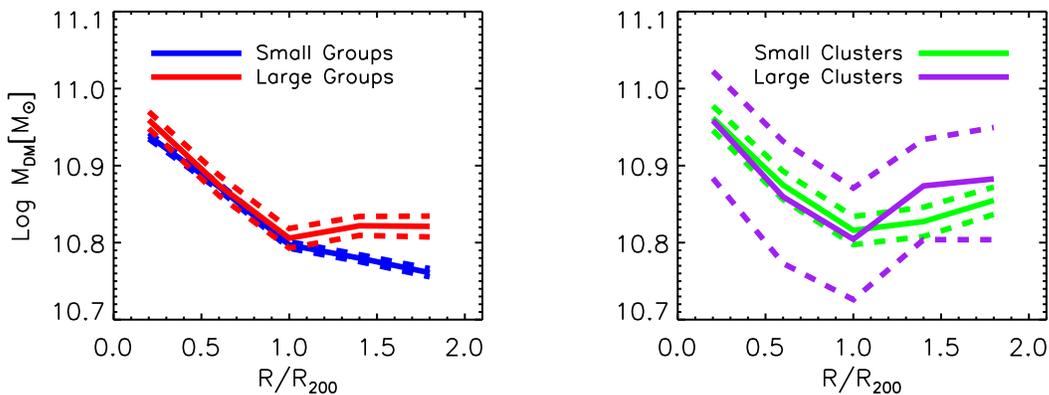}  % 0.75
\caption{Mean dark matter mass in subhaloes of galaxies with mass larger than $10^{10} \, M_{\odot} h^{-1}$
in the same samples of haloes shown in the bottom panels of Figure \ref{fig:mass_segr}. Due to the presence 
of "orphan" galaxies in our model (see text), this population of galaxies are missed here.}
\label{fig:mass_segr_dm}
\end{center}
\end{figure*}

\begin{table}
\caption{Slopes, zero-points and their errors obtained with a linear fit, $\log M_* = a \cdot (R/R_{200}) +b$,
done out to the virial radius, for all samples shown in the upper panels of Figure \ref{fig:mass_segr}.}
\begin{center}
\begin{tabular}{llllll}
\hline
Sample & a & $\sigma_a$ & b & $\sigma_b$ \\
\hline
Small Groups   & -0.678 & 0.012 & 10.797 & 0.005 \\
Large Groups   & -0.483 & 0.023 & 10.728 & 0.011 \\
Small Clusters & -0.339 & 0.021 & 10.665 & 0.011 \\
Large Clusters & -0.227 & 0.045 & 10.681 & 0.023 \\
\hline
\end{tabular}
\end{center}
\label{tab:tab2}
\end{table}

In the top panels of Figure \ref{fig:mass_segr} we plot the mean stellar mass of galaxies as a function of radial distance 
considering the mass at redshift $z=0$ (solid lines) and at the time of accretion \footnote{The time of accretion of a given 
galaxy is defined as the last time the galaxy is central.} (dotted lines) for our 4 samples of haloes (different colours), 
considering all satellite galaxies (explicitly excluding centrals) with stellar mass larger than $10^{10} M_{\odot}$ belonging 
to the FOF-group. We find a clear mass segregation up to the virial radius, with more massive galaxies preferentially located 
in the innermost regions, close to the centre. To better quantify the strength of the segregation we compute a linear fit 
out to the virial radius for all samples and show the result in Table \ref{tab:tab2}. The zero-point ($b$) decreases from small 
groups to big clusters of about 0.2 dex, while the slope ($a$) flattens significantly from -0.678 in small groups to -0.227 in 
big clusters. Moreover, none of the slopes are consistent with zero even within $3\cdot \sigma$.

Interestingly, at $r \sim R_{200}$ we find an upturn, and the mean stellar mass starts to increase again up to $2\cdot R_{200}$, 
from small groups to small clusters. As shown by the plot, the result still holds when considering the stellar mass at the time 
of accretion. In this case the trend is the same, but the mean stellar mass at each radial distance is slightly lower. This means 
that galaxies have had time to grow in mass by redshift $z=0$, but also that the effects of stellar stripping and star formation 
do not drive trends in mass segregation, suggesting that the main driver is of dynamical nature. We compare our results with 
results of V14 given by \citealt{dlb}'s model (black symbols in the top panels of Figure \ref{fig:mass_segr}). Despite these points
are shifted-low with respect to ours, they clearly show the same trend and a hint of upturn in the external regions in small and 
large groups, while they suggest neither mass segregation within the virial radius nor upturn in large clusters. 

\begin{figure*} 
\begin{center}
\includegraphics[scale=.75]{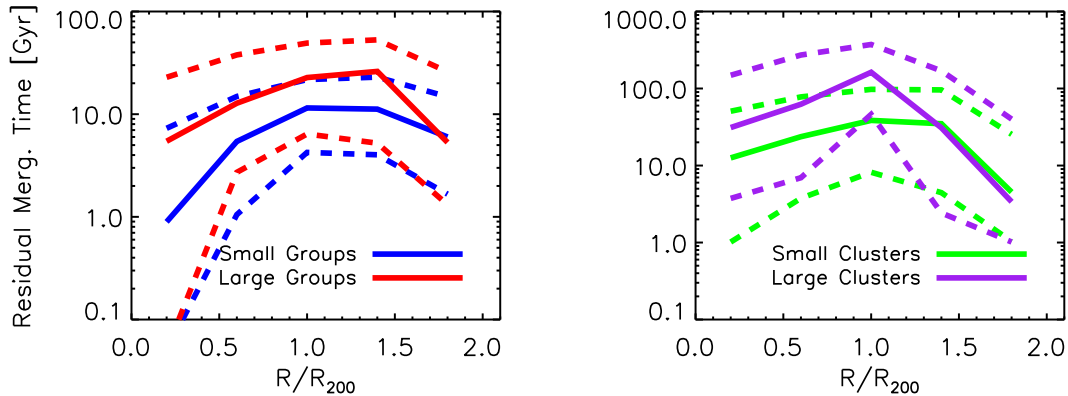} % 0.75
\caption{Residual merging time at $z=0$ (derived as the dynamical friction timescale minus the time 
elapsed since accretion) as a function of their radial distance from the halo centre. All galaxies  within 
a sphere centred on the halo centre and having a radius equal to $2\cdot R_{200}$ are considered.}
\label{fig:tdyn_all}
\end{center}
\end{figure*}

In the bottom panels of Figure \ref{fig:mass_segr} we show the same information plotted in the top panels, now considering all 
galaxies within a sphere centred in the centre of the halo and having a radius equal to $2\cdot R_{200}$. This procedure results 
in a sample including all of those galaxies in the top panels of Figure \ref{fig:mass_segr}, as well as all galaxies within the 
sphere that are not gravitationally bound to the group/cluster, and results in a sample that is more comparable to observational
samples. A comparison between the top and bottom panels indicates that adding galaxies not belonging to the FOF-group, but within 
$2\cdot R_{200}$, attenuates the steepness of the stellar mass-radial distance relation beyond $R_{200}$ in groups, and basically 
flattens it in clusters. 

In Figure \ref{fig:mass_segr_dm} we show the dark matter mass (DM) in subhaloes associated with galaxies with stellar mass larger than 
$10^{10} \, M_{\odot} h^{-1}$, for our samples of haloes. In our semi-analytic model we keep following the so-called "orphan-galaxies",
i.e. those galaxies that have lost their subhalo after accretion in a larger system. In Figure \ref{fig:mass_segr_dm} we take the 
samples of galaxies considered in the bottom panels of Figure \ref{fig:mass_segr}, without the orphan-galaxies. The trend of mass with 
radius shown by DM is the same found for stellar mass, a clear mass segregation up to the virial radius and an upturn beyond it, in all 
samples except for the small groups. This suggests that mass segregation might be driven mostly by dark matter rather than by physical 
processes involving only stellar mass. 

The prediction of our model is then consistent with observations that find evidence for mass segregation in groups and 
clusters (\citealt{vandenbosch08b,presotto12,balogh14,roberts15}), at least up to the virial radius. The upturn beyond the 
virial radius disappears on cluster scales, particularly when the sample includes non-FOF galaxies. We repeated the same analysis 
including low-mass galaxies (down to $10^{9} M_{\odot}$). The inclusion of low-mass galaxies only changes the mean stellar mass 
at each radial distance, but does not significantly change the overall trend. 

%%%%%%%%%%%%%%%%%%%%%%%%%%%%%%%%%%%%%%%%%%%%%%%%%%%%%%%%%%%%%%%%%%%%%%%%%%%%%%%%%%%%%%%%%%%%%%%%%%%%%%%%%%%%%%%%%%%%%%%%
\section{Discussion}
\label{sec:discussion}
Mass segregation in evolved objects is generally believed to be a consequence of dynamical friction, which causes more massive 
galaxies to fall to the centre of haloes more quickly. Nevertheless, the assembly history of haloes, as well as other physical processes 
taking place in dense environments, might play an important role. Galaxies can grow in stellar mass via star formation or mergers, 
and the rate of the growth could depend on their stellar mass. In addition, tidal forces in groups and clusters cause stellar stripping. 
Although dynamical friction brings massive galaxies towards the centre, this is also the environment in which tidal stripping gets stronger 
(\citealt{myself2}). It is possible that tidal stripping could be strong enough to mix the population of galaxies and no mass 
segregation would be present. We investigated on this switching-off our stellar stripping and merger channels for the formation of the 
intra-cluster light, and found similar results as those reported in Figure \ref{fig:mass_segr} (plots not shown), meaning that stellar 
stripping is not enough for mixing the galaxy distribution. This is confirmed by V14's data, that are predictions of \citealt{dlb} 
models, which does not have any prescription for stellar stripping.

As seen in Section \ref{sec:results}, the top panels of Figure \ref{fig:mass_segr} show that beyond $R_{200}$ the stellar mass-radial 
distance relation changes slope until it flattens in clusters. When considering all galaxies (bottom panels of Figure \ref{fig:mass_segr}) 
in the sphere centred on the centre of the halo, the slope flattens at lower halo mass, in large groups. This may be due to pre-processing 
and accretion of new galaxies by the FOF-group, which could reduce the strength of "anti-mass segregation" found outside the virial radius, 
that progressively weakens from small groups to large clusters. This is supported by the result found in Figure \ref{fig:mass_segr_dm}
for DM subhaloes. Accretion of smaller haloes and mergers of objects with comparable mass during the assembly of each halo have different 
consequences depending on the virial mass of the object that they are infalling onto or merging with. Large groups and clusters form 
later, and this may lead to the upturn due to recent mergers of DM haloes. In terms of stellar mass we do not find an upturn on cluster scales 
due to the non-linear relation between DM and stellar mass. This means that, in our model, the stellar mass growth in massive haloes is not as 
steep as the DM mass growth, and that the star formation efficiency is higher in low-mass haloes with respect to larger ones. This is consistent 
with the predicted stellar mass function, from which it is clear that semi-analytic models find more low-mass galaxies than observed
(see, e.g., \citealt{qiguo} and references therein).

We investigate the relationship between dynamical friction and halo mass in Figure \ref{fig:tdyn_all}, where we plot the residual 
merging time at $z=0$ versus the radial distance, for the same samples of haloes and considering all galaxies in the sphere centred 
on the centre of the halo and having a radius equal to $2\cdot R_{200}$, as done in the bottom panels of Figure \ref{fig:mass_segr}. 
The residual merging time is simply derived as the dynamical friction timescale minus the time elapsed since accretion. The dynamical 
friction timescale has been evaluated at the time of accretion and following \citealt{jiang08} formula (eqn. 5 of their paper).
Two interesting features arise from this plot. First, galaxies that are closer to the centre are also those with the smallest residual 
merging time and, overall, the relations fit well with the behaviour shown in the bottom panels of Figure \ref{fig:mass_segr}. Second, 
the residual merging times increase from groups to clusters. This means that mass segregation in groups can be explained by invoking 
dynamical friction, which has enough time to bring massive galaxies in the innermost regions of the group. This still applies in 
clusters, where the typical dynamical friction timescale is long enough to distribute galaxies according to their mass, but on average 
much longer than in groups (even longer than a Hubble time). 

We are aware that the "anti-mass segregation" we find beyond the virial radius is not observed (e.g. \citealt{roberts15}). There 
might be several reasons for that, including contamination of foreground and background galaxies. We do not have any contamination 
of such objects by construction. On the other hand, when we include unbound galaxies as done in the bottom panels 
of Figure \ref{fig:mass_segr} instead of FOF galaxies only (as done in the top panels of the same figure), the relation considerably 
flattens. As argued above, this might be due to recent mergers between DM haloes, which could reduce the strength of 
”anti-mass segregation” found outside the virial radius. The upturn beyond $R_{200}$ is remarkably sharp in groups, and appears 
sensitive to the inclusion of unbound galaxies. A full explanation of the upturn deserves a more detailed analysis,
which must take into account the merging histories of the haloes considered. We plan to address this point in a future paper. 

%%%%%%%%%%%%%%%%%%%%%%%%%%%%%%%%%%%%%%%%%%%%%%%%%%%%%%%%%%%%%%%%%%%%%%%%%%%%%%%%%%%%%%%%%%%%%%%%%%%%%%%%%%%%%%%%%%%%%%%%
\section{Conclusion}
\label{sec:conclusion}
In this letter we focus on the mass segregation in simulated groups and clusters coupled with a state-of-art semi-analytic 
model of galaxy formation that accounts for the physics of baryons. 

We find a non-negligible mass segregation in groups and clusters up to the virial radius, and the level of segregation is 
statistically significant at all halo masses. The strength of the segregation is found to be a function of halo mass, such
that is weaker at higher halo mass. This can be explained by invoking dynamical friction, which brings more massive galaxies faster 
towards the innermost regions of the halo. The mass segregation trends are insensitive to both the inclusion of low-mass 
galaxies in the sample (down to stellar mass $M_* = 10^9 \, M_{\odot}$) and stellar stripping.

Moreover, beyond the virial radius we find an "anti-mass segregation" in groups that progressively weakens in clusters. The 
upturn beyond the virial radius highlights that these galaxies are intrinsically more massive than those in the inner regions. 
Galaxies in the outskirts have been recently accreted and thus have been centrals for a longer time. They then had more time and 
more chance to grow in mass, due to cooling and enhanced star formation. Considering all galaxies in a sphere centred on the 
centre of the halo and having a radius equal to $2\cdot R_{200}$ (i.e. not only those belonging to the FOF-group), the trend 
beyond the virial radius weakens, approaching observational findings.

Interestingly, mass segregation is found also by looking at the dark matter in subhaloes, with similar features shown by the stellar 
mass. This result, along with the fact that neither stellar stripping nor star formation (nor both together) shape mass segregation 
after accretion, points out that the main driver of the radial distribution of galaxies has a dynamical nature. We showed that dynamical 
friction is the most likely candidate, but, in order to quantify its relative contribution in shaping mass segregation  and confirm the 
nature of the upturns beyond the virial radius, a more detailed analysis is needed.

%%%%%%%%%%%%%%%%%%%%%%%%%%%%%%%%%%%%%%%%%%%%%%%%%%%%%%%%%%%%%%%%%%%%%%%%%%%%%%%%
\section*{Acknowledgements}
The authors thank Benedetta Vulcani for providing us her data, Alessio Romeo for 
helpful comments, and the anonymous referee for improving the content of this letter.
EC and XK acknowledge financial support by NSF of Jiangsu Province (No. BK20140050), 
973 program (No. 2015CB857003, 2013CB834900), the NSFC (No. 11333008), and
the Strategic Priority Research Program the emergence of cosmological structures” of the 
CAS (No. XDB09010403). EC is also funded by Chinese Academy of Sciences 
Presidents' International Fellowship Initiative, Grant No.2015PM054.

\label{lastpage}

\bibliographystyle{mn2e}
\bibliography{biblio}

\end{document}